\documentclass[fleqn,usenatbib]{mnras}

\usepackage{newtxtext,newtxmath}

\usepackage[T1]{fontenc}

\DeclareRobustCommand{\VAN}[3]{#2}
\let\VANthebibliography\thebibliography
\def\thebibliography{\DeclareRobustCommand{\VAN}[3]{##3}\VANthebibliography}

\usepackage{graphicx}	
\usepackage{amsmath}	
\usepackage{bm} 
\usepackage{xcolor}
\usepackage{hyperref} 
\hypersetup{colorlinks=true, citecolor=blue, urlcolor=blue, linkcolor=blue}

\def\tsig{\tilde{\sigma}}
\def\romanI{\uppercase\expandafter{\romannumeral1}}
\def\romanII{\uppercase\expandafter{\romannumeral2}}

\title{Force chains bias the dynamic response to impacts in rubble-pile asteroids}

\author[C. Huang et al.]{
Chenyang Huang$^{1,2}$
Yang Yu$^{1}$\thanks{E-mail: yuyang.thu@gmail.com}
Peter R. King$^{2}$
Bin Cheng$^{3}$
Raphael Blumenfeld$^{4}$\thanks{E-mail: rbb11@cam.ac.uk}
\\
$^{1}$School of Aeronautic Science and Engineering, Beihang University, Beijing 100191, China\\
$^{2}$Department of Earth Science and Engineering, Imperial College London, London SW7 2BP, UK\\
$^{3}$School of Aerospace Engineering, Tsinghua University, Beijing 100084, China\\
$^{4}$Gonville \& Caius College, University of Cambridge, Cambridge CB2 1TA, UK}

\date{Accepted XXX. Received YYY; in original form ZZZ}

\pubyear{\the\year{}}

\begin{document}
\label{firstpage}
\pagerange{\pageref{firstpage}--\pageref{lastpage}}
\maketitle

\begin{abstract}
The impact response of rubble-pile asteroids is essential for both elucidating their formation and evolution history and evaluating the efficacy of impact defense strategies. Although state-of-the-art numerical simulations have allowed for the replication of many macroscopic impact characteristics consistent with observations, the understanding of dynamics and response mechanisms within rubble-pile structures remains incomplete and requires further in-depth investigation. Such understanding is critical for assessing the effects and safety of impact defense missions.
The loose structure of rubble-pile asteroids affects inhomogeneous internal stress propagation via inherent force chains, which may lead to structural fracturing. We demonstrate this phenomenon here, using a proof-of-principle two-dimensional model of granular aggregates.
We find that the velocity response front to impact disturbances preferentially propagates along pre-existing force chains, with particles not in chains responding more slowly. The sites within the response zone where high dynamic stresses manifest are strongly correlated with these initial force chains, and the damages that result are predominantly located within areas enclosed by these chains. 
The strong correlation between pre-existing force chains and dynamic response is independent of the location, magnitude, direction of the disturbance velocity, or the aggregate's particle size distribution. 
All evidence suggests that the core reasons for this propagation preference lie in the structural heterogeneity of granular aggregates and the resulting differences in mechanical wave propagation.
This investigation provides guidance for future research aimed at quantitatively assessing fragmentation risks based on the statistical properties of force chains. 
\end{abstract}




\section{Introduction}
The moniker rubble pile is typically applied to small bodies in the Solar System that measure between 200 m and $\sim$10 km in diameter~\citep{walsh2018rubble}. Decades of observations, in situ detections, and various theoretical, numerical, and experimental studies have led to a consensus about rubble-pile asteroids: they are aggregates reformed by the self-gravity of fragments from the collisional disintegration of primitive asteroids, with significant porosity between their irregularly shaped components~\citep{johansen2015new,Klahr_Delbo_Gerbig_2022}. Over the past two decades, exploration/deflection missions to asteroids Itokawa, Ryugu, Bennu, and Didymos-Dimorphos system have provided unprecedented details about rubble piles~\citep{fujiwara2006rubble,watanabe2019hayabusa2,lauretta2019unexpected,daly2023successful}. We can directly observe their surface regolith layers with particles ranging from micrometers to meters, impact craters, and unique topographical features~\citep{sugita2019geomorphology,walsh2019craters}, and gain insights into their physical and chemical properties~\citep{yada2022preliminary,walsh2022near,robin2024mechanical}, although direct measurements of the internal structures of rubble piles are still lacking.

Impact events serve as pivotal exogenous factors influencing the formation and evolution of rubble-pile asteroids. The topographical and geological features of these celestial bodies often unveil clues consistent with non-destructive impact scenarios. Beyond prominent impact craters, diverse phenomena—ranging from crater erosion and obliteration~\citep{thomas2005seismic,hofmann2017small} to the development and proliferation of fissures~\citep{prockter2002surface,massironi2012geological} and the size segregation of regolith particles~\citep{matsumura2014brazil,maurel2016numerical}—have been linked to the dynamics of stress wave propagation triggered by impacts.
To gain deeper insights into asteroid properties and to mitigate potential impact hazards, humans initiated in situ impact experiments~\citep{arakawa2020artificial,daly2023successful}. A landmark event occurred in September 2022 when NASA's Double Asteroid Redirection Test (DART) mission successfully modified the orbit of Dimorphos using a kinetic impactor~\citep{cheng2023momentum}, which is a full-scale demonstration of kinetic impact defense technology. Post-impact observations revealed that while the DART mission generated significant ejecta and a persistent tail, the target body, Dimorphos, retained most of its mass~\citep{li2023ejecta}, herein referred to as the ``remaining asteroid''. However, the precise responses on the surface and internal structure of the remaining rubble-pile asteroid remain elusive.
ESA's Hera mission, launched in October 2024, is designed to conduct a thorough investigation of the DART impact's consequences~\citep{michel2022esa}.
Therefore, understanding impact-driven processes is essential not only for decoding the natural evolutionary paths of these bodies but also for supporting the design and demonstration of future planetary defense missions.

For non-catastrophic impact events such as NASA's DART mission, the dynamic response induced by the residual kinetic energy propagating within the remaining asteroid warrants close attention. Here the residual kinetic energy is equal to the impactor's kinetic energy minus: the instantaneous dissipated energy during the hypervelocity impact, the kinetic energy carried away by ejecta, and the kinetic energy increment carried by the translational and rotational changes of the remaining rubble pile. This insight aids in assessing the impact effects and safety, while also enhancing our understanding of the evolution of rubble-pile asteroids driven by stress wave propagation.
More broadly, this prompts inquiry into the dynamics of impact disturbance propagation through heterogeneous and inelastic granular clusters and the ensuing responses of such clusters.
Advancements in numerical simulations have significantly enriched our comprehension. \cite{jiao2024sph} employed a hybrid Smoothed Particle Hydrodynamics (SPH) and Discrete Element Method (DEM) framework to meticulously model the entire dynamic evolution of Dimorphos during the DART impact, from the initial shock wave through to the propagation of stress wave, providing insights into resurfacing processes during the subsequent self-gravitating evolution phase. \cite{liu2024impact}, building on data derived from terrestrial impact tests concerning the proportion of residual energy~\citep{walker2013momentum}, utilized the DEM approach to simulate the propagation of residual kinetic energy within the remaining asteroid, elucidating aspects such as the energy spreading range, crater dimensions, and surface depressions resultant from resurfacing. \cite{tancredi2023lofting} adopted a comparable strategy to simulate material ejection at low speeds and the surface vibrations induced by seismic waves.
These studies have shed light on the macroscopic evolution of the remaining rubble-pile asteroids driven by the propagation of residual energy. Yet, the specific propagation patterns of such disturbances at mesoscopic and microscopic scales, and the associated internal structural responses, remain largely uncharted territories.

In this work, we developed a proof-of-principle numerical model to gain insight into the dynamic response of granular aggregates to impact disturbances. Through detailed particle-level stress analysis, we identified inherent force chain structures within the clusters. We observed that the velocity response to impact disturbances preferentially follows pre-existing force chains. Locations of elevated dynamic stresses, and the resultant contact failures and structure damages, exhibit a strong correlation with these initial force chains. The propagation preference of impact disturbances, as dictated by these force chain structures, appears impervious to variations in the magnitude and direction of the disturbance velocity, the specific location of the disturbed particle, or the overall particle size distribution of the cluster. This work illustrates the critical role of force chain structures in the post-impact response of granular aggregates, using a non-average method at the mesoscopic scale for the first time.

\section{Methods}

\subsection{Granular aggregate structure characterization}
\label{sec:structcharact} 
\begin{figure}
    \centering
    \includegraphics[width=0.5\columnwidth]{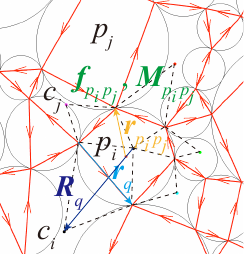}
    \caption{Characterization of the granular aggregate structure and the inter-particle contact interactions. Cells' polygon edges are depicted in red with arrows (clockwise vector group). The black dashed-line quadrilaterals are the quadrons that belong to, and represent the area of, particle $p_i$. $\bm{r}_q$ and $\bm{R}_q$ are the diagonals of a quadron. $\bm{f}_{p_ip_j}$ and $\bm{M}_{p_ip_j}$ respectively denote the force and torque moment that particle $p_j$ applies to particle $p_i$ at their contact position. $\bm{r}_{p_ip_j}$ is the vector extending from the center of mass of $p_i$ to the $p_i$-$p_j$ contact position.}
    \label{fig_cell}
\end{figure}

We use the quadron and cell method~\citep{ball2002stress,blumenfeld2004stresses} to quantify disordered structures of 2D self-gravitating granular assemblies.
The method is based on constructing the contact network: the network's nodes are the contact points, edges are the lines connecting the nodes around each particle, and cells are the polygons representing the smallest voids surrounded by particles in contact, as shown in Fig.~\ref{fig_cell}. The edges are directional, rotating around each particle in the anti-clockwise direction. Thus, the cell polygons representing the inter-particle voids are composed of clockwise vectors. Each edge is a vector characterized uniquely by particle $p$ and cell $c$ that $p$ shares. For example, in Fig.~\ref{fig_cell} the edge vector $\bm{r}_{q}$ is determined by particle $p_i$ and cell $c_i$. 

Next, a dual network is constructed by extending a vector $\bm{R}_{q}$ from each particle's centroid to the centroids of the cells surrounding it. The centroids of particles and cells are defined as the mean positions of the contacts around them. The vectors $\bm{r}_{q}$ and $\bm{R}_{q}$ are dual and cross one another except for very non-convex particle shapes, which is not the case in this study. Each such pair of vectors forms a quadrilateral, of which they are the diagonals. This quadrilateral has been named quadron~\citep{blumenfeld2004stresses}. The shape of each quadron is quantified by the tensor $C_q = \bm{r}_q \otimes \bm{R}_q$. The anti-symmetric part of this tensor provides its area, $A_q$,
\begin{equation}
\frac{1}{2} \left(C_q - C_q^T\right) = A_q \left(
\begin{matrix} 
0 & 1 \\
-1 & 0 \\
\end{matrix} 
\right)\ ,
\end{equation}
in which $C_q^T$ is the transpose of $C_q$. 

For assemblies of convex cell shapes, quadrons tessellate perfectly 2D granular assemblies, $A_\mathrm{system}=\sum_{q} A_q$. Each particle's area is determined by its force-carrying contacts and the cells it shares. The area associated with particle $p$ is 
\begin{equation}
A_{p} = \sum_{q\in p} A_q \, 
\end{equation}
with $q$ running over the quadrons that belong to $p$. For example, the area of particle $p_i$ is the sum of the areas of the 5 quadrons represented by the black dashed lines in Fig.~\ref{fig_cell}.

Occurrence of star-like and very tortuous cells may lead to overlapping neighboring quadrons, which could lead to errors in estimating the total system area and the particle area~\citep{blumenfeld2003granular,ciamarra2007comment,blumenfeld2007blumenfeld}. For the granular aggregates in our simulations, the total area of such quadrons was smaller than $0.22\%$. Therefore, simply handling the overlapping quadron areas does not introduce significant errors in the calculation of particle areas.

\subsection{Particle-scale stress tensor}

In 2D granular aggregates, the inter-particle contact interactions include normal force, tangential force, and rolling torque (the contact force models will be introduced in Section~2.3). In Fig.~\ref{fig_cell}, the resultant force exerted by particle $p_j$ on particle $p_i$ is denoted as $\bm{f}_{p_ip_j}$, the torque as $\bm{M}_{p_ip_j}$ (i.e., the rolling torque), and the contact position between the two particles is represented by the vector $\bm{r}_{p_ip_j}$ originating from the center of mass of particle $p_i$.

We derived particle-scale stress tensor according to the definition of the stress tensor for Representative Element Volume in granular media~\citep{nicot2013definition,yan2019definition}, which are derived from the first principles and considered the dynamic effects related to particle rotation. The stress tensor of particle $p_i$ is denoted as 

\begin{equation}
\begin{split}
&\sigma_{p_i} = \\
&\frac{1}{A_{p_i}}\left[\begin{array}{ll}
\sum\limits_{p_j} f_{p_ip_j,x} r_{p_ip_j,x}+J_{p_i}\omega_{p_i}^2 & \sum\limits_{p_j} f_{p_ip_j,x} r_{p_ip_j,y}+J_{p_i}\dot{\omega}_{p_i} \\
\sum\limits_{p_j} f_{p_ip_j,y} r_{p_ip_j,x}-J_{p_i}\dot{\omega}_{p_i} & \sum\limits_{p_j} f_{p_ip_j,y} r_{p_ip_j,y}+J_{p_i}\omega_{p_i}^2
\end{array}\right] \ ,
\end{split}
\end{equation}

\noindent where $p_j$ are particles contacting with $p_i$, $\bm{f}_{p_ip_j}$ the force that $p_j$ exerts on $p_i$, $\bm{r}_{p_ip_j}$ the position vector to the contact between $p_i$ and $p_j$ (see Fig.~\ref{fig_cell}), $J_{p_i}$ the moment of inertia of $p_i$, $\omega_{p_i}$ its angular velocity, $J_{p_i}\dot{\omega}_{p_i}$ the total resultant moment on $p_i$, and $A_{p_i}$ the area associated with particle $p_i$. 


\subsection{N-body Soft-Sphere DEM code \texttt{DEMBody}}
In this work, we simulated the formation of self-gravitating rubble-pile asteroids (granular aggregates) and tracked their responses to impact disturbances using the N-body soft-sphere discrete element method (SSDEM) code \texttt{DEMBody}, which is developed for granular dynamics simulation in planetary science and geophysics~\citep{cheng2018collision,cheng2019numerical,cheng2021reconstructing,huang2024cumulation}. 

In this code, the nonlinear contact forces between two contacting spheres are calculated based on the Hertzian normal force model and Mindlin–Deresiewicz tangential force model~\citep{somfai2005elastic}. The normal and tangential forces are described as the following formula:

\begin{equation}
\begin{aligned}
F_n & =\frac{4}{3} \sqrt{r_{\mathrm{eff}}} E_{\mathrm{eff}}^* \delta_n^{3 / 2}-\gamma_n u_n \ ,\\
F_s & =\min \left(\mu\left|F_n\right|, 8 \sqrt{r_{\mathrm{eff}} \delta_n} G_{\mathrm{eff}}^* \delta_s-\gamma_s u_s\right) \ .
\end{aligned}
\end{equation}

\noindent $\delta_n$ and $\delta_s$ ($u_n$ and $u_s$) denote the normal and tangential mutual displacements (velocities) between two contacting particles, and the mutual normal overlap $\delta_n$ is typically smaller than $1\%$ of the radii. The damping coefficient $\gamma_n$ and $\gamma_s$ correlate with the restitution coefficient $\epsilon_n$ and $\epsilon_s$~\citep{wada2006numerical}, and in this work $\epsilon_n=\epsilon_s=0.4$, falling within the reasonable range for rock materials. The effective radius $r_\mathrm{eff}$, effective Young's modulus $E_{\mathrm{eff}}^*$, and effective shear modulus $G_{\mathrm{eff}}^*$ are derived from the harmonic mean of the properties of contacting particles $i$ and $j$: $r_{\mathrm{eff}}^*=r_i r_j /\left(r_i+r_j\right)$, $\frac{1}{E_{\text {eff }}^*}=\frac{1-\nu_i^2}{E_i}+\frac{1-\nu_j^2}{E_j}$, $\frac{1}{G_{\text {eff }}^*}=\frac{2\left(2-\nu_i\right)\left(1+\nu_i\right)}{E_i}+\frac{2\left(2-\nu_j\right)\left(1+\nu_j\right)}{E_j}$.

The non-sphericity of the actual grains introduces resistance to rolling and twisting motions. \texttt{DEMBody} incorporated a spring-dashpot rotational model~\citep{jiang2015novel} to mimic the elastic and plastic characteristics. In this work, particles only experience rolling resistance within 2D aggregates (monolayers of spheres on the same plane). Here we give the formula of rolling torque:  

\begin{equation}
\begin{aligned}
& M_r=\min \left(0.525\left|F_n\right| \beta r_{\mathrm{eff}}, 0.25\left(\beta r_{\mathrm{eff}}\right)^2\left(k_n \delta_r-\gamma_n \omega_r\right)\right) \ .\\
\end{aligned}
\end{equation}

\noindent $\delta_r$ and $\omega_r$ represent the relative rolling angular displacement and velocity between two contacting particles, respectively. The spring stiffness coefficient $k_n=\frac{4}{3}\sqrt{r_{\mathrm{eff}}\delta_n}E_{\mathrm{eff}}^*$. The shape parameter $\beta$ characterizes the statistical deviation of a real particle from a sphere, which introduces the effects of non-spherical particles on the mechanical properties. The combination of friction coefficient $\mu$ and shape parameter $\beta$ can control the macroscopic characteristics of cohesionless particles, such as the internal friction angle~\citep{cheng2023measuring}, which further influences the porosity of granular aggregates. 
\section{Results}
\subsection{Simulation setup}
The rubble-pile asteroids are simplified and modeled as particle aggregates in DEM simulations. To observe the propagation of impact disturbances within heterogeneous internal structures clearly, we employed a 2D rubble-pile model, i.e., monolayer particles are confined to the same plane. The initial particle aggregate was constructed by placing a four-particle nucleus at the origin and $N\approx3000$ dispersed particles inside an annulus surrounding the nucleus. The particles collapse under mutual gravity toward the nucleus. The nucleus is designed to ensure that the particles in the annular region are attracted to the center more quickly and stabilize there. Particle motion was damped by a viscous term to dissipate the kinetic energy on the approach to a static state. The granular aggregates were considered static when the total kinetic energy fell below $10^{-6}$ times the total initial gravitational potential energy and the contact network did not change anymore.

Most of our investigations were carried out using an equal-numbers bi-disperse particle size distribution (PSD): $d_1=5$ m and $d_2=10$ m, with or without several surface boulders ($d=20, 40, 50$ m). We also ran simulations with a power-law PSD, $P(>d)\sim d^{-3}$, with different $d$ ranges ($5~\mathrm{m}\leq d\leq 10~\mathrm{m}$ or $5~\mathrm{m}\leq d\leq 15~\mathrm{m}$).
The cumulative size frequency distributions with power-law exponents ranging from $-2.65$ to $-3.5$ have been repeatedly observed within specific size ranges on the surfaces of rubble-pile asteroids.~\citep{mazrouei2014block,michikami2019boulder,walsh2019craters}. 
The porosity/filling fraction of the post-accretion granular aggregates was controlled by the particle friction properties. We ultimately employed particle aggregates with area filling fractions ranging from 0.75 to 0.79. The particle properties and the contact mechanics parameters used to generate such aggregates are shown in Table~\ref{table1}. The mechanical stability of the granular aggregates was ascertained by checking that the bulk mean coordination numbers, excluding rattlers, were slightly larger than 3~\citep{ball2002stress}. 

\begin{table}
\centering
\caption{Particle characteristics and contact mechanics parameters in our simulations.}
\label{table1}
\begin{tabular}{lcl}
\hline
Parameter & Symbol & Value\\
\hline
Young modulus & $E$ & $5\times10^9$ Pa\\
Poisson's ratio	& $\nu$ & $0.3$\\
Friction coefficient & $\mu$ & $1.0$ \\
Restitution coefficient	& $\epsilon$ & $0.4$ \\
Shape parameter	& $\beta$ & $0.8$ \\
Material density & $\rho$ & $3.4\times10^3$ kg/m$^3$ \\
\hline
\end{tabular}
\end{table}

To investigate the propagation patterns of residual kinetic energy within the remaining rubble-pile asteroid, we consider an elementary process: how a velocity increment in particles within a cluster disturbs the entire cluster structure. This elementary process is simulated by applying initial velocity to particles or boulders on the surface of clusters.
We also applied impulses of velocity perturbations to particles that belong to force chain structures {\it inside} the cluster, which are useful for elucidating the role of force chains in the preferential propagation of impact disturbances. In the following, we call these perturbations also impact, although they are not physically possible. 
We tested the magnitudes of velocity impulses ranging from $5\times10^{-3}$ m/s to 5 m/s, corresponding to the momenta from $8.90\times10^{3}$ kg$\cdot$m/s to $8.90\times10^{6}$ kg$\cdot$m/s. 
We mainly focused on the propagation patterns of impact disturbances in scenarios where the overall particle cluster structure is without significant damage. Significant damage here refers to conditions where more than half of the particles are scattered, rearranged, or even result in the breakup of the cluster.
The low limit of the velocity magnitude was chosen to enable observations of preferred propagation directions of particle speed and stress, which get washed out by extended damage on more violent impacts, like the up-limit magnitude. Additionally, the range of disturbing velocities we tested encompasses the velocity magnitudes of residual asteroid components following DART impact, as simulated by~\cite{jiao2024sph,liu2024impact}.

For generality, we nondimensionalize all physical quantities in the following analyses. We use non-dimensional lengths, time, velocities, and stress components, respectively scaled by: 
\begin{equation}
l_0\equiv\bar{d}\ ;\ 
t_0\equiv\sqrt{\frac{l_0}{\bar{g}}}\ ;\ 
v_0\equiv\frac{l_0}{t_0}\ ;\ 
\sigma_0\equiv\frac{\bar{m}}{t_0^2} \ .
\end{equation}
Thus, $t\to\tau=t/t_0$, $\bm{v}\to\bm{u}=\bm{v}/v_0$, $\sigma\to\tsig=\sigma/\sigma_0$, and lengths are in units of $l_0$. Here, $\bar{d}$ is the mean particle diameter, $\bar{g}=(\mathrm{G}/N)\sum_{i}\sum_{j\neq i}m_j/\left|\bm{r}_j-\bm{r}_i\right|^2$ is the pre-impact gravitational acceleration magnitude, averaged over all the cluster particles, and $\bar{m}$ is the mean particle mass. 

\subsection{The propagation patterns of impact disturbance}
\begin{figure*}
    \centering
    \includegraphics[width=1\linewidth]{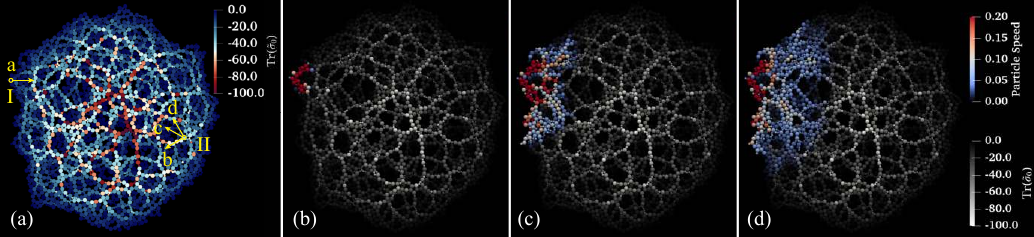}
    \caption{(a) Pre-impact bi-disperse granular aggregates, colored by the initial stress level Tr($\tsig_0$). Impact disturbances a-d (yellow arrows) are applied at particles {\romanI} and {\romanII} (yellow circles, $M=1.78\bar{m}$). (b)-(d) Particle speed propagation at $\tau=2.65\times10^{-4}$, $1.59\times10^{-3}$, $2.92\times10^{-3}$ in the case of Impact ``a'' with $u_\mathrm{impact}=10$. The particle speed layer in blue-red is superimposed on the layer denoting pre-impact stress Tr($\tsig_0$) in gray scale. In the speed layer, particles with speed $u\lesssim 0.01$ are transparent.}
    \label{fig_ptspeed}
\end{figure*}

A typical example of the non-uniform stress distribution in a self-gravitating bi-disperse particle aggregate is shown in Fig.~\ref{fig_ptspeed}(a). The level of compressive stress in particles is measured by the trace of the particle stress tensor $\mathrm{Tr}(\tsig_0)$, through which the inherent force chains in a cluster --- specifically, the chain-like structures formed by particles with relatively higher compressive stresses, shown in reddish hues in Fig.~\ref{fig_ptspeed}(a) --- are identified clearly. Note that the compressive stress is indicated with a negative sign, so the higher the absolute value of Tr($\tsig_0$), the greater the compressive stress experienced by the particles. 
In the bi-disperse packing, the occurrence probability of $\left|\mathrm{Tr}(\tsig_0)\right|$ decays exponentially, and the probability that $\left|\mathrm{Tr}(\tsig_0)\right|>100$ is 0.0073. In Fig.~\ref{fig_ptspeed}(a), particles with Tr($\tsig_0$)$\leq -100$ are colored in the same deep red to enhance the visual representation of the Tr($\tsig_0$) distribution across the entire cluster. 
Similarly, in the particle coloring strategy presented below, the range that the color bar covers is set to better reflect the differences between particles and the distribution characteristics of the corresponding physical quantities within the particle cluster. Values that exceed the range of the color bar correspond to the same colors as the maximum or minimum values displayed on the color bar.

In this section, we will introduce the propagation patterns of impact disturbance consistent across multiple simulations from the perspectives of particle speed, particle stress, microscopic failure events, and microscopic fractures. Here, we present the simulation results where particle {\romanI} ($d=10$~m) is subjected to an impact disturbance in direction ``a'' (see Fig.~\ref{fig_ptspeed}(a)), with a perturbation velocity $u_\mathrm{impact}=10$ (corresponding to $v_\mathrm{impact}=0.05$~m/s). The dimensionless impact momentum is $17.80$.

Figure \ref{fig_ptspeed}(b)-(d) exhibit the particle speed propagation after disturbance, where the particle speed layer in blue-red is superimposed on the pre-impact stress layer in gray scale to show the correlation. The underlying Tr($\tsig_0$) layer is colored in gray scale, allowing particles under weak stress to blend with the black background, thereby highlighting the structure of force chains. In the top speed layer, particles with speeds $u\lesssim 0.01$ are set transparent, ensuring they do not obscure the display of the stress layer. We observed that the particle speeds preferentially propagate along the force chains. At the intersections of force chains, particle speeds circumvent the black areas in the stress layer, which are regions occupied by weak-stress particles, and prefer to propagate along the force chains. After the speed front passes through, particles in these weak stress areas also receive very low-magnitude velocity perturbations caused by their surrounding force-chain particles. The propagating pattern of particle speed is also available in the supplemental video ``bidisp\_u10.mp4''.

\begin{figure*}
    \centering
    \includegraphics[width=0.8\linewidth]{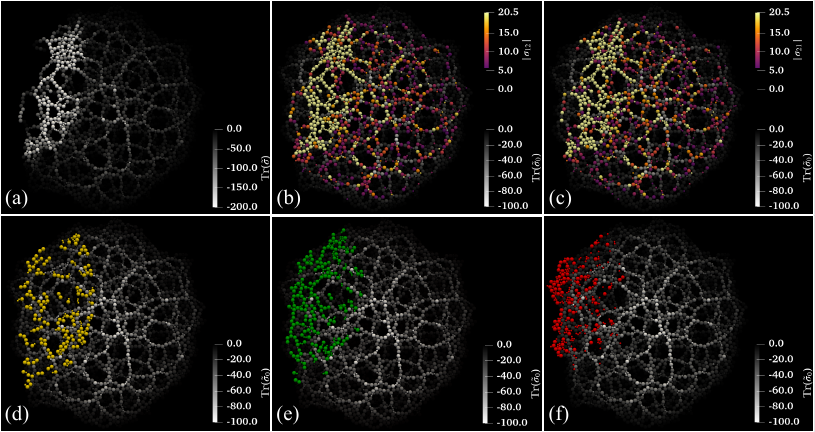}
    \caption{Propagation of particle stress and microscopic failure events at $\tau=2.92\times10^{-3}$ in the case of Impact ``a'' with $u_\mathrm{impact}=10$. (a) Dynamic normal stress level Tr($\tsig$). (b) The layer in purple-yellow indicating the absolute value of dynamic shear stress component $\lvert\tsig_{12}\rvert$ is superimposed on the layer denoting pre-impact stress Tr($\tsig_0$) in gray scale. In the shear stress layer, only particles with $\lvert\tsig_{12}\rvert$ values in the top $30\%$ are colored using a gradient from purple to yellow. (c) Dynamic shear stress component $\lvert\tsig_{21}\rvert$, with the same rendering strategy as panel (b). (d) and (e) Contact pairs experiencing sliding and rolling failures are marked in yellow and green respectively, and the underlying layer still represents the Tr($\tsig_0$). (f) Particles with coordinate number=0 are rendered in red.}
    \label{fig_stress_event}
\end{figure*}

The dynamic stress analyses show that particle stresses propagate faster and are more pronounced along the initial force chains. 
Figure~\ref{fig_stress_event}(a) illustrates the compressive stress Tr($\tsig$) propagation at $\tau=2.92\times10^{-3}$, by which time the propagation front has reached about $1/3$ of the cluster. At this moment, particles under higher compressive stresses, depicted in white, are primarily distributed along the initial force chain structure. For the particles near the disturbance partilce {\romanI}, the compression waves have already passed through, releasing elastic potential energy, and they are temporarily in a scattered or weak-stress state.
All the panels in Fig.~\ref{fig_stress_event} correspond to the moment shown in Fig.~\ref{fig_ptspeed}(d). 
We also rendered particles within the top $30\%$ of shear stress in a purple-yellow gradient as shown in Fig.~\ref{fig_stress_event}(b) and (c). The color bar's upper and lower limits correspond to the absolute values of the shear stress component ($\lvert\tsig_{12}\rvert$ and $\lvert\tsig_{21}\rvert$) in the top $10\%$ and the top $30\%$, respectively. In the prevailing range of impact disturbance propagation, there is a noticeable overlap between particles experiencing the highest shear stress and those initially in the force chains.

The preferential propagation of shear stresses can also induce failures of particle contacts. We documented the occurrence of sliding and rolling failure events between particles. Sliding failure occurs when the tangential force between two particles exceeds the maximum static friction that the contact surface can provide, resulting in relative sliding between the particles. Similarly, according to the rolling deformation friction model, rolling failure occurs when the rolling torque between particles exceeds a certain threshold, leading to relative rolling between them. In Fig.~\ref{fig_stress_event}(d) and (e), contact pairs that undergo sliding and rolling failures are marked in yellow and green, respectively. We found that failure events frequently occur in weak-stress particles that are in contact with particles from the initial force chains. The distribution of the scattered particles, i.e., particles whose coordinate number became 0, also supports this point (see Fig.~\ref{fig_stress_event}(f)). Besides the particles near the disturbed particle {\romanI}, other scattered particles are primarily distributed within the ``black'' cavities formed by the initial force chains.

\subsection{The role of force chains}
\begin{figure*}
    \centering
    \includegraphics[width=0.6\linewidth]{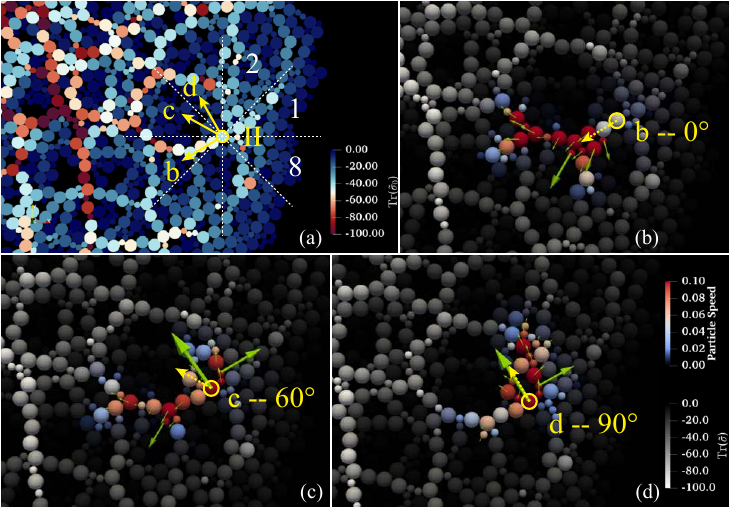}
    \caption{(a) A local magnification of the bi-disperse cluster around particle {\romanII}. The cluster is divided into 8 sectors centered at particle {\romanII}, labeled by white digits. (b)-(d) Particle speed propagation at $\tau=6.63\times 10^{-4}$ after velocity disturbances ($u_\mathrm{impact}=1$) applied on particle {\romanII} at $0^\circ$, $60^\circ$ and $90^\circ$ to the local force chain respectively, corresponding to impact ``b'', ``c'', and ``d'' marked in panel (a). The green arrows denote the orientations of particle velocities at the current moment, whose lengths are proportional to the magnitudes of particle velocities. The yellow dashed arrows indicate the directions of perturbation velocity, not scaled according to the velocity magnitude. 
    }
    \label{fig_internaldisturb}
\end{figure*}

Section 3.2 demonstrates that force chains determine the preferred propagation paths for impact disturbances. In this section, we select the particles belonging to the force chains from the interior of the cluster to isolate the role of force chains. Here we show the perturbation velocities ($u_\mathrm{impact}=1$) are applied on particle {\romanII}  ($d=10$~m) in directions of $0^\circ$, $60^\circ$, and $90^\circ$ relative to the local force chain, corresponding to impact disturbances ``b'', ``c'', and ``d'' respectively, as shown in Fig.~\ref{fig_internaldisturb}(a). The local force chain mentioned here refers to a series of particles aligned in the direction of arrow ``b'' in Fig.~\ref{fig_internaldisturb}(a).

As a control group, following the application of impact disturbance ``b'', the transmission of particle velocities predominantly occurs along the force chains. Upon encountering a force chain bifurcation, should the antecedent particle have a velocity component directed towards this divergence, there will be a significant propagation along that branched path (see Fig.~\ref{fig_internaldisturb}(b)). 
In impact test ``c'', we found that particles with high speeds are still prevalent along the force chain we focused on (oriented $60^\circ$ anticlockwise from the disturbance direction) and there is little further propagation directly along the disturbance direction, as shown in Fig.~\ref{fig_internaldisturb}(c). As perturbing velocity disseminates along the force chain, particle {\romanII} undergoes a cumulative force and torque from adjacent particles, prompting a clockwise rotation in its velocity direction. Subsequently, pronounced velocity transmission manifests on the force chain branches within the clockwise sector relative to the original perturbation direction (the yellow dashed arrow). 
Even when impact disturbances are applied orthogonal to the force chains, velocity propagation along these chains is evident, such as disturbance ``d'' in Fig.~\ref{fig_internaldisturb}(d). This tangential spread is driven partly by the tangential forces and rolling moments between the disturbed particle and its contacts, and partly by the velocity direction shift of the initially impacted particle through interactions with its neighbors, initiating an increase from zero in the velocity component along the force chain we focused. 
Overall, we noted rapid and pronounced propagation of particle velocities on force chains in non-disturbed directions. Especially when there are no continuous particle-contact pathways in the direction of disturbance, nearby force chains become the preferred routes for propagation. The heterogeneous structure of particle clusters makes this preference for force chain propagation a general phenomenon.

Beyond specific observations at the particle level, we also introduced statistical indicators to quantitatively describe the anisotropy in disturbance propagation caused by the force chain structures.
We define a measure of the mean propagating distance of particle speed response,
\begin{equation}
    D=\frac{\sum_{i=1}^{N} H\left(\left|\bm{u}_i\right|-\left|\bm{u}_\mathrm{th}\right|\right) \left|\bm{r}_i-\bm{r}_\mathrm{impact}\right|}{R_\mathrm{g}\sum_{i=1}^{N} H\left(\left|\bm{u}_i\right|-\left|\bm{u}_\mathrm{th}\right|\right)} \ ,
\end{equation}
in which $\bm{r}_\mathrm{impact}$ and $\bm{r}_i$ are the position vectors of the impacted particle and particle $i$, respectively, with the origin at the aggregate's center of mass, $\bm{u}_i$ the velocity of particle $i$, $\left|\bm{u}_\mathrm{th}\right|=\alpha\left|\bm{u}_\mathrm{impact}\right|$ a velocity threshold, and 
$H(x)$ is the Heaviside step function.
$D$ is normalized by the aggregate's radius of gyration $R_\mathrm{g}$. The following definition of $R_\mathrm{g}$ is commonly used to measure the size of particle agglomerates, 
\begin{equation}
    R_\mathrm{g} = \sqrt{\frac{\sum_{i=1}^N m_i r_i^2}{\sum_{i=1}^N m_i}} \ ,
\end{equation}
where the individual mass elements $m_i$ are located at distances $\bm{r}_i$ from the center of mass \citep{wurm1998experiments}.
We find that while $D$ depends somewhat on the choice of $\alpha$, the conclusions are not.

\begin{figure}
    \centering
    \includegraphics[width=1\linewidth]{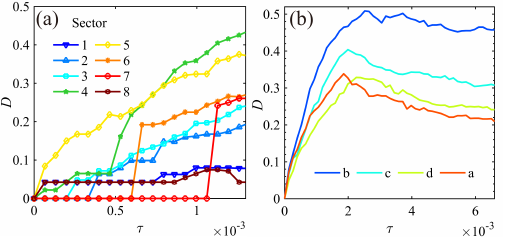}
    \caption{(a) The evolution of $D$ in the 8 sectors around particle {\romanII} in impact test ``c'', with $u_\mathrm{impact}=1$ and $u_\mathrm{th}=0.01$. 
    (b) The evolution of the total value of $D$ in impact tests a-d. $u_\mathrm{impact}=1$, and $u_\mathrm{th}=0.01$.}
    \label{fig_Dt}
\end{figure}

In Fig.~\ref{fig_Dt}(a), we show the mean propagating distance of dynamic response, $D$, in the different sectors which are marked in Fig.~\ref{fig_internaldisturb}(a) in the case of impact disturbance ``c'' with $u_\mathrm{impact}=1$.
Particle {\romanII} is on a force chain that extends into sector 5, then curves into sector 4, and continues into sector 3. Figure~\ref{fig_Dt}(a) confirms that $D$ indeed propagates fastest into sector 5. At $\tau\approx5\times10^{-4}$, $D$ increases dramatically in sector 4, corresponding to the moment when the stress response, carried by the chain, entered sector 4. In contrast, $D$ remains low in sector 6 until this time owing to the lack of force chains directly connected to particle {\romanII}, and in sectors 1, 7, and 8, which are opposite to the impact direction. Later, $D$ increases in sector 6 and then in sector 7, reflecting the eventual arrival of the stress response to these sectors along the chain paths rather than directly through the medium.

We also investigated the temporal evolution of the mean propagating distance $D$ across the cluster under assorted impact perturbations. Figure~\ref{fig_Dt}(b) delineates the $D-\tau$ trajectories for perturbations directed along ``b'', ``c'', and ``d'' applied to particle {\romanII}, which are situated on the cluster's intrinsic force chain. Perturbation along arrow ``a'' on the weak-stress surface particle {\romanI} (marked in Fig.~\ref{fig_ptspeed}(a)) served as a reference.
We observed that propagation is ``the most efficient'' when the disturbance is aligned with the force chain (case ``b'') and ``the least efficient'' when it is perpendicular (case ``d'') in the comparison. Here, ``more efficient'' refers to a scenario in which the slope of $D$ is steeper during the rise phase (indicating faster advancement of the dynamic response) and it maintains a higher level during the plateau phase (indicating further propagation of the impact response).
Consistent with the above, we see in Fig.~\ref{fig_Dt}(b) that the initial response ($\tau\lesssim2\times10^{-3}$) to the impact disturbance on a weak-stress surface particle (case ``a'') propagates much more slowly than on an initial-force-chain particle and along the force chain (case ``b''). The slope for case ``a'' initially approaches that of case ``c'' and subsequently falls behind case ``c'', but remains steeper than case ``d'' throughout. This occurs because in case ``a'', the weak-stress surface particle {\romanII}, once disturbed, initially transmits the perturbation through particles that it contacts to nearby force chain structures, which are oriented at certain angles relative to the direction ``a''.

\subsection{The effect of the perturbation velocity magnitude}
In this section, we assessed the effect of the perturbation velocity magnitude on the propagation patterns of impact disturbances. The results indicate that within the range of perturbation velocities tested, $1\leq u_\mathrm{impact}\leq10^3$, the propagation into undamaged regions is preferentially along the meandering force chains regardless of $u_\mathrm{impact}$. At higher perturbation velocities, such as $u_\mathrm{impact}=10^3$, a significant number of particles behind the propagation front are scattered, and those near the disturbed particle show noticeable displacements, resulting in damage to the original aggregate structures.

\begin{figure}
    \centering
    \includegraphics[width=1\linewidth]{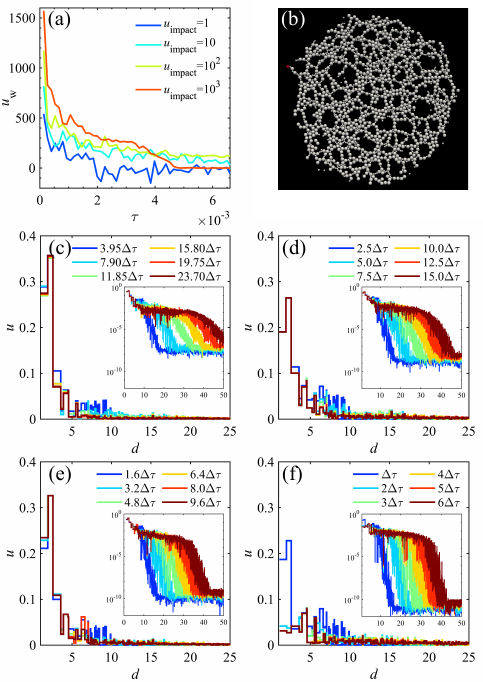}
    \caption{(a) The mean propagation speed of particle velocity responses under perturbations of varying magnitudes. $u_\mathrm{th}=0.01$. (b) Particles with initial compressive stress $\lvert$Tr($\tsig_0$)$\rvert$ larger than the median value are indicated. The red one denotes the location of particle {\romanI}. (c)-(f) Particle speed distribution when particle {\romanI} is applied impact disturbances along direction ``a'' (marked in Fig.~\ref{fig_ptspeed}(a)). Panels (c)-(f) correspond to the disturbance magnitudes $u_\mathrm{impact}=1, 10, 10^2, 10^3$, respectively. The stair plots represent the distribution of particle speeds within strong stress structures shown in panel (b) as a function of their distance from the disturbed particle. Particle speed $u$ in $y$ axis is normalized by $u_\mathrm{impact}$. The distance $d$ in $x$ axis is scaled by mean particle size $\bar{d}$. Different colors indicate different moments in time. $\Delta\tau=2.65\times10^{-4}$. The insets in panels (c)-(f) show semi-log plots of the same particle speed distributions as the main plots, but in larger distance scales.}
    \label{fig_speedspectrum}
\end{figure}

We define a smoothed mean disturbance propagation speed,
\begin{align}
    u_{\mathrm{w}}(\tau)=\frac{D(\tau)-D(\tau-T)}{T} \ ,
\end{align}
with $T=1.33\times10^{-4}=50$ computational time steps. The evolution of $u_{\mathrm{w}}$ at different disturbance magnitudes is shown in Fig.~\ref{fig_speedspectrum}(a), in which $u_\mathrm{th}=0.01$ for the determination of $D$.
The response propagation speed diminishes swiftly as the propagation advances, with the trajectories for various perturbation velocities converging and eventually exhibiting minor oscillations at lower levels. Notably, the trajectory for $u_\mathrm{impact} = 10^3$ stabilizes at zero, indicating that the velocities of all particles have exceeded $u_\mathrm{th}$, rendering $D$ constant. The comparison in Fig.~\ref{fig_speedspectrum}(a) illustrates that the influence of disturbance velocity magnitude on propagation speed within the granular aggregate structure exhibits nonlinear characteristics.

We determined the force chain structures in every realization as particles carrying compressive stress $\lvert$Tr($\tsig_0$)$\rvert$ higher than the median value (see Fig.~\ref{fig_speedspectrum}(b)). For the bi-disperse aggregates, $\lvert$Tr($\tsig_0$)$\rvert_\mathrm{median}/\lvert$Tr($\tsig_0$)$\rvert_\mathrm{max}=0.054$. The higher this threshold ratio the more sparse the force chain network.
We next analyze the spatial distribution of speed response of these force chain particles under various perturbation velocity magnitudes, $u_\mathrm{impact}=1, 10, 10^2, 10^3$, as shown in Fig.~\ref{fig_speedspectrum}(c) to (f) respectively. We compare the moments when the particle speed responses caused by different $u_\mathrm{impact}$ propagate to almost the same distance, which referred to the propagation time power law $\tau\sim u_\mathrm{impact}^{-0.2}$ from the impact simulations of one-dimensional sphere chains (see Supplemental Materials). In Fig.~\ref{fig_speedspectrum}(c) to (f), the stair plots of the same color represent moments when the propagation distances of velocity responses are very close. 
At these moments, the particle velocities near the disturbed particle ($d<6$) almost do not decay over time in Fig.~\ref{fig_speedspectrum}(c), (d), and (e), while the velocity decay after $\Delta\tau$ in Fig.~\ref{fig_speedspectrum}(f) is obvious. This is due to the high-level perturbation at $u_\mathrm{impact}=10^3$, where particles near the disturbed particle experience significant displacements, altering their contact relationships and resulting in new collisions.
Besides, we observed the speed at these moments decays exponentially with distance at the front of the response regions in the insets of Fig.~\ref{fig_speedspectrum}(c)-(f), which show the semi-log plots of the same distributions as the main plots. And the larger the perturbation velocity, the steeper the decay.

\begin{figure*}
    \centering
    \includegraphics[width=1\linewidth]{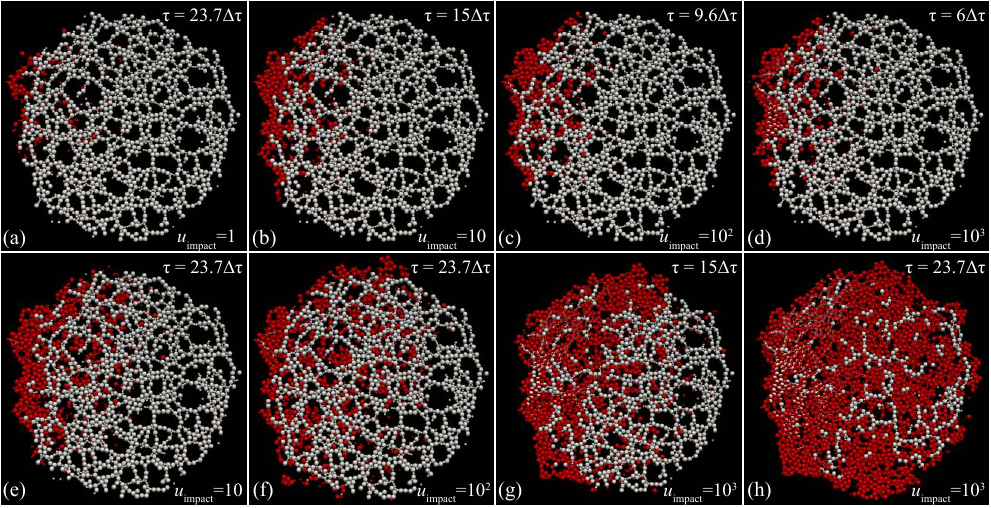}
    \caption{Distribution of scattered particles (particles with no contacts) in red and pre-impact strong-stress particles in white. Scattered particles are represented on the upper layer of the diagram. Panels (a)-(d) show the dissociated particles when the velocity response propagates the same distance under different disturbance magnitudes, $u_\mathrm{impact}=1, 10, 10^2, 10^3$, respectively. The impact disturbance is applied on particle {\romanI} along arrow ``a''. These four moments correspond to those of the reddish-brown stair plot curves in Fig.~\ref{fig_speedspectrum}. Panels (a), (e), (f), and (h) exhibit damage distribution after identical propagation time for $u_\mathrm{impact}=1, 10, 10^2, 10^3$, respectively. Panels (e) and (f) display the late-stage evolution for $u_\mathrm{impact}=10^3$.}
    \label{fig_cn0distri}
\end{figure*}

Impact disturbances also generate scattered particles, which become dissociated from the force-carrying structure. These particles at different impacts and moments are shown in red and appear in the upper layers, while initial force chain particles are white in Fig.~\ref{fig_cn0distri}. If scattered particles overlap with force chain particles, the red will cover the white.
Figure~\ref{fig_cn0distri}(a) to (d) display the distribution of scattered particles at the moments corresponding to the reddish-brown stair plots in Fig.~\ref{fig_speedspectrum}(c) to (f), respectively, at which the propagation distances of dynamic responses under the four perturbation velocities are very close. Figure~\ref{fig_cn0distri}(a), (e), (f), and (h) compare the scattered particles after identical propagation time for different disturbance magnitudes. At $\tau=23.7\Delta\tau$, the disturbance at $u_\mathrm{impact}=10^3$ has penetrated the entire cluster. 
Except for Fig.~\ref{fig_speedspectrum}(h), we observed that, besides particles in the shallow surface layers near the perturbation point, the other internal scattered particles are mainly distributed within the cavities enclosed by the initial force chain structures, consistent with Section 3.2. And most of those are adjacent to the initial force chains. Thus, the particles carrying the highest loads are not particularly damaged regardless of $u_\mathrm{impact}$ (including the early stage of $u_\mathrm{impact}=10^3$ case), but the damage zones increase with $u_\mathrm{impact}$. 
Figure~\ref{fig_cn0distri}(e) and (f) show the scattered particles in the late stages at high perturbation $u_\mathrm{impact}=10^3$. As the dynamic response penetrates the entire cluster, even the force chains are damaged at this magnitude.
Over sufficiently long periods, these dissociated particles will reconvene and coalesce under self-gravity, forming new force chain structures —-- a topic for future research. Notably, many red particles do not completely cover the white particles, especially at $u_\mathrm{impact}=10^3$, indicating particles that have undergone significant displacement.

\begin{figure}
    \centering
    \includegraphics[width=0.75\linewidth]{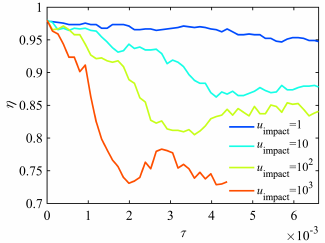}
    \caption{The proportion of particles within the top $30\%$ of dynamic shear stress values that also fall within the top $50\%$ of initial compressive stress values.}
    \label{fig_overlapratio}
\end{figure}

To test our conjecture that the high correlation between the highest shear stress and the pre-existing force chains is independent of the perturbation magnitudes, we identified two subsets of particles: those that fall within the top $50\%$ of the pre-impact compressive stress ($S_1$) and those that experience the highest $30\%$ dynamic shear stress ($S_2$). Figure~\ref{fig_overlapratio} displays the fraction of particles that fall within $S_2$ and also overlap with $S_1$, i.e., $\eta=N(S_1\cap S_2)/N(S_2)$, as a function of time for different $u_\mathrm{impact}$.  
 The plot for $u_\mathrm{impact}=10^3$ should be regarded as a lower bound because of the deterioration of the force chains in the late stages of the evolution. Nevertheless, even so, the overlap between the two subsets is higher than $72\%$. For mild disturbance, $u_\mathrm{impact}=1$, the overlap even consistently exceeds $95\%$.

\subsection{The effect of the composition of particle clusters}
All the aforementioned tests of impact disturbance are conducted using the bi-disperse granular aggregates. This section will introduce the disturbance simulations on bi-disperse packing with surface boulders and multi-disperse packing whose particle size distribution (PSD) follows a power law. The results support that the propagation preference of impact disturbance shows insensitivity to the composition of particle clusters, at least within the scope of this study. 

\begin{figure*}
    \centering
    \includegraphics[width=1\linewidth]{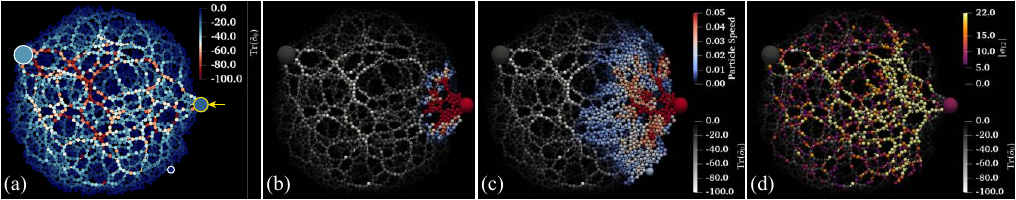}
    \caption{(a) Pre-impact bi-disperse granular aggregates with three boulders (circled), colored by the initial stress level Tr($\tsig_0$). Impact disturbance toward the center of mass of the aggregate is applied on the boulder ($M=85.57\bar{m}$) circled in yellow with magnitude $u_\mathrm{impact}=0.21$.
    (b) and (c) Particle speed propagation at $\tau=1.37\times10^{-3}, 4.38\times10^{-3}$, respectively. The particle speed layer in blue-red is superimposed on the layer denoting pre-impact stress Tr($\tsig_0$) in gray scale. In the speed layer, particles with speed $u\lesssim 0.0025$ are transparent. 
    (d) The layer in purple-yellow indicating the absolute value of dynamic shear stress component $\lvert\tsig_{12}\rvert$ is superimposed on the layer denoting pre-impact stress Tr($\tsig_0$) in gray scale at $\tau=4.38\times10^{-3}$. In the shear stress layer, only particles with $\lvert\tsig_{12}\rvert$ values in the top $30\%$ are colored using a gradient from purple to yellow.}
    \label{fig_bdr}
\end{figure*}

Surface of real rubble-pile asteroids is scattered with massive boulders, giving rise to the force chains \citep{cheng2021reconstructing} whose strengths depend on the boulders' masses. This kind of force chains generated from surface particles is lacking in the bi-disperse packing. To assess the effects posed by these chains, we added to the static bi-disperse particle clusters three surface boulders ($d=20, 40, 50$ m), which are released with zero velocity from a distance comparable to the original cluster size. Figure~\ref{fig_bdr}(a) shows that strong force chains formed only under the two more massive boulders. We ran several simulations with different disturbed boulders and various magnitudes of disturbance velocity. Here we show the typical results through a case that the impact disturbance toward the aggregate's center of mass is applied to the 40 m boulder with $u_\mathrm{impact}=0.21$. The dimensionless impact momentum is $17.80$, the same as the impact disturbance applied on particle {\romanI} in Section 3.2. Capture of particle speed propagation in Fig.~\ref{fig_bdr}(b) and (c) illustrates that the speed fronts tend to propagate preferentially along the initial force chains, and particles in the weak stress area behind the speed front are disturbed by nearby force chain particles that have already responded, resulting in a delayed velocity response. Fig.~\ref{fig_bdr}(d) provides evidence that the higher dynamic shear stress is strongly correlated with the initial force chain structure. 
From the direction of disturbance propagation, the next disturbed layer beneath the boulder comprises several smaller particles in contact with it. Combining the examples from Section 3.2, where a single particle is disturbed, we can conclude that the propagation pattern of impact disturbance responses is independent of whether the disturbance is initiated at a single point or multiple points. Thus, our findings from most scenarios, which investigate the elementary processes of responses within rubble-pile structures due to residual kinetic energy, can be extended to real scenarios where particles in the same region are disturbed simultaneously.

\begin{figure*}
    \centering
    \includegraphics[width=1\linewidth]{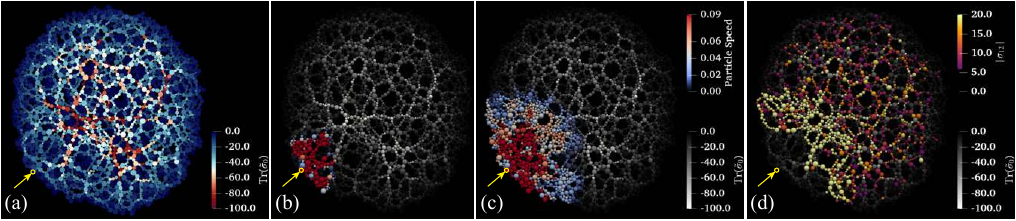}
    \caption{(a) Pre-impact multi-disperse granular aggregates whose particle size follows a power law with exponent -3, colored by the initial stress level Tr($\tsig_0$). Impact disturbance toward the center of mass of the aggregate is applied on a surface particle ($M=3.84\bar{m}$) circled in yellow with magnitude $u_\mathrm{impact}=4.64$. (b) and (c) Particle speed propagation at $\tau=8.00\times10^{-4}, 2.67\times10^{-3}$, respectively. (d) Dynamic shear stress component $\lvert\tsig_{12}\rvert$ propagation at $\tau=2.67\times10^{-3}$. The coloring strategies for particle speed and shear stress are the same as in Fig.~\ref{fig_bdr}.}
    \label{fig_multidisp}
\end{figure*}

To ascertain that our results are not limited to bi-disperse particle clusters, we ran similar simulations with a power-law PSD, $P(>d)\sim d^{-3}$. We chose $5\leq d\leq15$ m to ensure the same cluster mass as in the bi-disperse case for a fair comparison. Fig.~\ref{fig_multidisp} depicts an example of the response to an impact on a surface particle with $u_\mathrm{impact}=4.64$. The impact momentum is $17.80$, the same as the particle {\romanI} case in Section 3.2 and the 40 m boulder case in this Section. In all the simulations of these systems, an example of which is shown in Fig.~\ref{fig_multidisp}, we observed the same preferential and enhanced propagation of the dynamic response in the force chain direction. The weak stress areas are circumvented by the advancing particle velocity front and exhibit delayed responses, as shown in Fig.~\ref{fig_multidisp}(b) and (c). In some cases, the high-speed front is even perpendicular to the impact direction as it follows the chain direction. Fig.~\ref{fig_multidisp}(d) also reveals that whether in the wavefront or in undisturbed structures, high shear stress locations coincide with pre-impact force chains.
Further simulations of aggregates with the same power-law exponent $-3$ and the same particle size range as the bi-disperse range ($5\leq d\leq10$ m), which are less massive, also support the aforementioned conclusions. 

\section{Discussion and Conclusion}
In this study, we observed that the leading edge of velocity response due to impact disturbances propagates along initial force chains of the self-gravitating aggregates. Meanwhile, the velocity response in weak stress areas is delayed, typically triggered by secondary disturbances caused by surrounding responded force chain particles. The mechanism for this propagation preference can be explained from two aspects: firstly, weak stress areas often have structural flaws, lacking sufficient particle contacts or effective pathways for mechanical wave propagation; secondly, the speed of sound in granular systems correlates positively with the degree of compression~\citep{lherminier2014revealing}, which is also verified through our one-dimensional impact disturbance tests (see Supplemental Materials). In our one-dimensional sphere chain tests, we found that larger initial inter-particle embedding leads to faster peak response propagation of particle velocity, with embedding ranges covering $10^{-9}\bar{d}$ to $10^{-5}\bar{d}$ in the initial static particle clusters.
In the propagation process of disturbance responses, particles in the response area, typically along several particles' widths in the direction of propagation, exhibit high dynamic stresses (compressive components and shear components) closely related to the initial force chains. The correlation between dynamic compressive stress and initial force chains is also demonstrated in the photoelastic experiments by~\cite{owens2011sound}, which shows that the highest amplitude sounds travel along the strongest force chains. The amplitude of sound in experiments correlates positively with the compressive stress we observed in simulations. It is always challenging to measure the propagation patterns of shear stress in granular media by experiments, but high-fidelity DEM simulations allow us to provide detailed observations, which is crucial as high shear stress poses a great risk of structural fracture and damage. On a granular scale, high shear stress leads to contact failures between particles, and we observed that sliding and rolling failures primarily occur between force chain particles and their non-chain contacts. We found the scattered particles, besides those near the disturbed surface particles, are mainly distributed within the cavities formed by the initial force chains.

In summary, this work explores the propagation patterns of residual kinetic energy within the remaining rubble-pile asteroids posterior to the impact and ejection processes. A 2D proof-of-principle model is applied to explore the common mechanism of residual energy propagation. By using the 2D model we are allowed to calculate the particle-scale stress, which reveals the mesoscopic interaction mechanism among the self-gravitating aggregates in unprecedented accuracy. The results revealed the role of force chain structures in the preferential propagation of impact disturbances. The response to impact perturbations propagates preferentially along the initial force chains, which bear the majority of the load induced by the residual impact kinetic energy. It suggests the impact disturbance may travel further than expected along these invisible paths, rather than dissipated uniformly in all directions from the impact site. 
The velocity response front always propagates along pre-impact initial force chains, while non-chain particles respond with a delay. High dynamic stresses typically occur on initial force chain particles, where high dynamic shear stresses lead to contact failures and structure damages, thus these failures or damages are also closely related to initial force chains. 

Events like the DART mission, characterized by low specific energy $Q$ (the kinetic energy of projectile per unit target mass), cannot cause disintegration of the target. The range of perturbation velocities we applied was also limited to those that would not cause overall structural disruption. Within this range, we found that the propagation tendency of impact disturbance responses does not change with the magnitude of perturbation velocity when facing an undisturbed cluster ahead of the front; however, higher perturbation velocities can cause more particles behind the propagation front to scatter in a short time. We also found that the propagation pattern of impact disturbances is insensitive to different particle size distributions in cluster configurations. The large boulder disturbance scenarios represent more closely the residual kinetic energy propagation within the remaining asteroids after a real impact as they generate the multi-point or surface-triggered propagation, rather than a single-point trigger. Our analysis of dependency on propagation patterns also reiterates that the fundamental cause of non-uniform propagation of impact disturbances in particle clusters is the presence of force chains and defect structures. 
Therefore, it is reasonable to predict that the impact disturbance propagation patterns discovered in this work can qualitatively be generalized to 3D self-gravitating aggregates, in which force chains are inherently present. Additionally, it is important to note that the additional dimension might introduce variations in the energy dissipation rate. 
Future studies will focus on the propagation patterns of disturbances and their correlation with force chains inside rubble-pile asteroids, which may ultimately help us to infer the internal structure and compaction level of asteroids.

\section*{Acknowledgements}
C.H. is grateful for the hospitality of Imperial College London, where this work was carried out. C.H. acknowledges the international joint doctoral education fund of Beihang University. Y.Y. acknowledges the financial support provided by the National Natural Science Foundation of China Grants No. 12272018.

\section*{Data Availability} 
The data underlying this article will be shared at a reasonable request by the corresponding authors. The \texttt{DEMBody} code is available at \href{https://bin-cheng-thu.github.io/dembody-code/}{https://bin-cheng-thu.github.io/dembody-code/}.


\bibliographystyle{mnras}
\bibliography{ref} 





\bsp	
\label{lastpage}
\end{document}